# CODEVAL: IMPROVING STUDENT SUCCESS IN PROGRAMMING ASSIGNMENTS

**A. Agrawal, A. Jain, B. Reed**

*San José State University (USA)*

## Abstract

CodEval is a code evaluation tool that integrates with the Canvas Learning Management System to automatically evaluates students' work within a few minutes of the submission. This early feedback allows students to catch and correct problems in their submissions before their submission is graded and gives them a clear idea of the quality of their submission. CodEval handles the tedious aspects of grading, such as compiling and running tests, leaving graders more time to spend on the qualitative aspect of grading.

Before using CodEval, instructors would not have a clear view of the student's comprehension of the concept evaluated by the assignment until after the due date. CodeEval helps instructors identify and address the gaps in students' understanding and thus helps more students successfully complete the assignment.

We implemented CodEval using Python using the public Canvas API. Any instructor or grader for a Canvas course can use CodEval to automatically evaluate submissions for programming assignments. We developed a syntax to express requirements of submissions such as compilation parameters, inputs, outputs, command-line arguments, timeouts, exit codes, functions used, files generated, output validators, and more. We have made CodEval open source.

CodEval is an easy tool for students, graders, and instructors and seamlessly integrates with Canvas. We share our experience with using CodEval in two classes with a total of 90 students and multiple coding assignments.

## 1 INTRODUCTION

At San José State University, we have developed CodEval, a code evaluation tool integrated with the Canvas Learning Management System [1], that automatically evaluates students' work within a few minutes of the submission. We initially wanted to simply validate student submissions to programming assignments to ensure that they were submitting the correct code for the assignment and that their code was compiled, but we ended up developing a system that did extensive checking of the correctness of the submission including running test cases and implementation validation.

We use the Canvas Learning Management System from Instructure to manage assignments and track submissions and grades. Students zip up the code they want to submit and upload it through the assignment's page in Canvas. The grader can then download all the submissions from canvas, evaluate the submission, and post a grade with any relevant comments to the submission.

Graders evaluate submission by unzipping the code, compiling it, running it against test cases, and finally inspecting the code itself to verify that the correct techniques were used to implement the solution. Problematic submissions at each of those stages of grading make grading difficult and sometimes require contacting students to resubmit their assignments. Students sometimes purposely make problematic submissions to take advantage of a later submission. Since most of these stages tend to be tedious and time-consuming, it takes away from the grader's time to give quality feedback to students.

We wanted to design a system that would automate the tedious and manual aspects of the grading process to allow students to discover and correct submission problems before the grading started. We also wanted to have the compilation, testing, and initial code evaluation finished and summarized for the grader automatically before the grading process started. This allows graders to quickly review the summary and spend their time on the qualitative evaluation of the submission.

We finished the first implementation of CodEval in time for the Spring 2022 semester. We used it in our Operating Systems class where six substantial programming assignments were implemented using C. We also used it in an Algorithms class where programming assignments were implemented using Java. Results from these two classes achieved the goals we set out for. We caught all of the problematic submissions early so that the student had a chance to fix them themselves without involving the grader. Graders did not get involved in the actual compiling or running of the test cases. The feedback we collected from students was also positive.

## 2 REQUIREMENTS

We started out with some basic requirements:

- **simple integration with Canvas**: integrate with the existing submission workflow in Canvas and do not require the use of extra tooling for students and graders.
- **quick feedback**: allow students to receive feedback within minutes of uploading their submission. This allows students to correct any mistakes in the submission.
- **check compilation**: sometimes a submission will only compile on the author's own machine. This makes it impossible for a grader to evaluate their code accurately. Since the students use their own machines for development, the student may not even be aware of the problem. Providing feedback on problems detected after submission can make students aware of these issues.
- **tests that use exit code, stdin, stdout, stderr, provided files, and program arguments**: some of our assignments teach students the difference between stdin and stdout. Others test how to process program arguments. We need a testing framework that can validate these aspects of student submissions. Sometimes a student will misread assignment instructions and not realize it until the assignment has been graded. Providing early feedback on how a submission matches expectations allow students to catch and correct misunderstandings.
- **timeouts**: sometimes a student's submission will have errors such that the program never terminates or takes many minutes to run. These submissions would trip up previous ad-hoc testing by graders and were time-consuming to deal with.

Our initial implementation of CodEval met these requirements. Once we saw the feedback students were getting and the correctable problems that the students were still having, we found some additional enhancements to improve CodEval.

We realized we had other requirements we started the implementation:

1. **checking the use of specified functions**: most of our assignments required students to use certain APIs. Occasionally, a student would not read the instructions correctly and not use the correct APIs. We added a check that the submission used the required APIs. The grader still needed to verify that the API was used correctly but at least we knew that that student understood the API to use.
2. **validating generated files**: some of the assignments generate database-like files. Comparing an expected file to the generated file was not sufficient, so we needed a way to specify a validation program that instructors could supply to test generated files.
3. **checking concurrency and memory**: some of the assignments were to test students' understanding of threads and memory. We wanted to be able to accommodate instructor-provided programs to test concurrency and memory management. `valgrind` [2], is an example of a memory management validation program that we use.
4. **avoiding security issues using Docker**: normally graders will download submissions and compile and run them on their personal machines. When we thought about automatically downloading and running student submissions on the servers running CodEval, we realize that this opens up the server to accidental or malicious corruption by students. For example, a student could write a program that accidentally erases all the files on the server. Using Docker [3] can help up contain these problems by running submissions in a container that is rebuilt for each submission.

## 3 RELATED WORK

Automatic grading of programming assignments has been studied for many years, and similar to our project, most of these revolve around testing the assignment against a set of inputs and comparing the outputs. We looked at previous work with respect to the requirements we outlined previously.

Most of the tools support C++ and Java but lately, a number of tools have started supporting more languages [4]. Aldriye et Al. [5] and J Caiza et al., in [6], have compared and analyzed the latest automated grading tools on the market. They highlight the differences, the advantages, and the disadvantages between the systems and have paved the way for what can be the next steps for the automated grading tools. Several tools such as CourseMaker developed by Nottingham University[7], JavaBrat [8] developed by San Jose State University, WebCat [9], and RoboLift( an incremental work on WebCat) are used to grade assignments in programming languages Java, C++, Scala respectively.

As part of their evaluation metrics, CourseMaker [7] validates the student code against predefined test cases and checks typographic consistency (indentations, comments, layout of the file). WebCat [9] checks code correctness (how many tests pass), test completeness (check which parts of the code are actually executed), and test validity (test accurate-consistent with the assignment) whereas in JavaBrat[8] and Virtual Programming lab [10], grading is based on code correctness, which is determined by validating the output against the output from test cases. Some tools evaluate the code based only on test cases passed or failed, whereas others such as CourseMaker evaluate the completeness of test cases and additional criteria such as correctness of code design. CodEval uses the latter approach for validation and expands it further by introducing more comprehensive checks.

A. Gordillo [11], looked into the impact of automated grading technologies on students' perceptions of them as well as their performance. According to the findings, incorporating the automated evaluation tool into the course benefited students by increasing motivation, boosting the quality of their work, and improving their practical programming skills. As we will show, CodEval has similar results.

In [6], the author mentions that the next steps for the automated grading tools would be integration with LMSs since most of the tools such as CourseMaker [7], WebCat [9], and Marmoset [12] are standalone. For instance, users need to install the CourseMaker client on their system to use it. Although there exist some tools, such as Virtual Programming Lab, JAssess, and JavaBrat, that can be integrated with LMSs or provide plugins with platforms like Moodle, there is room for further improvement. CodEval aims to be scalable as it can be integrated with LMSs and also eliminates the requirement of installing specific software to validate code.

The grading tool by Magdeburg University [13], is not coupled with a given LMS. It considers three servers: the front-end, an LMS system, the spooler server, which controls the request, the submissions queues, and the back-end calls; which are the modules to evaluate a programming language. To communicate with the servers, XML-RPC (Remote Procedure Calls) has been used. The concept of making the automated grading tool independent of an LMS system is appealing. CodEval can be used independently with other LMSs as well. It does not require the use of any explicit procedure calls to communicate but uses the APIs of the LMS.CodEval also supports multiple languages and environments but leverages existing Docker containers to run rather than requiring separate servers to be set up.

Security is a crucial aspect of any software product. In automated grading tools, preserving the submissions and protecting the host server from any malicious attacks from any submission is critical. To protect against the consequences of malicious code, Marmoset [12] runs its J2EE web server and SQL database in a separate environment. To counteract authentication concerns and criminal actions, WebCat [9] has adjusted architectural designs. A more robust approach toward security is to use sandbox environments. CourseWork uses a sandbox environment to execute its code. Usage of containers has twofold advantages. They provide security to the host server and provide a customized and pristine environment to run the tests. CodEval uses Docker for its container environment, but other environments can be used by changing the configuration parameters.

Liu et Al. [14], have built a tool called AutoGrader that searches for semantically different execution paths between a student's submission and the reference implementation instead of generating feedback with failing test cases. The aim of this paper is to reduce the human effort required in generating test cases to automate the entire process of evaluating a submission based on semantics. CodEvals allows the integration of such tools for evaluation by specifying custom test commands in the evaluation specification.

# 4 SYSTEM DESIGN

As shown in Figure 1, CodEval runs on a local machine and communicates with Canvas using the Canvas REST API. This API allows CodEval to download submissions and add comments to the submissions. We also use the Canvas File service to store specification files and other support files for testing and evaluating the submissions. Canvas uses API tokens to restrict access to Canvas functionality so CodEval will only be able to access courses and content that the user running CodEval is able to access. This ensures that a student cannot access other students' submissions and specification files.

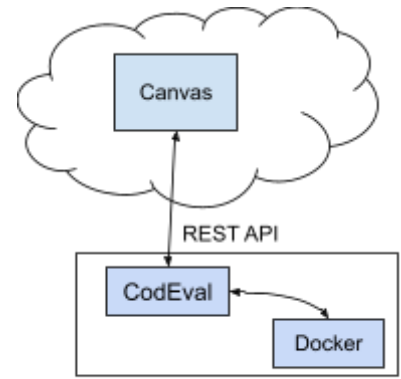

Figure 1: System architecture.

When CodEval has downloaded a submission and the specification files that describe how to evaluate the submission, it uses Docker to run the compilation and test cases in a container to isolate students' code from the system running the test. CodEval can also run the evaluations outside of Docker, but using a container like Docker is recommended to make sure that buggy or malicious submissions do not damage the machine running CodEval. After the submission has been evaluated, CodEval will add the results as a comment on the submission in Canvas using the Canvas REST API.

CodEval does not require any special Canvas configuration to work. It uses the REST API that instructors can enable by creating an access token. Instructors upload all the specification files and support files to Canvas so the instructor only needs to interact with the Canvas interface to create CodEval assignments. Once CodEval is running on a server using the provided tokens, instructors, students, and graders do everything through Canvas.

## 4.1. CodEval specification files

We designed the CodEval specifications files to be easy to write and read. We also wanted them to be flexible enough to accommodate both our simple programming assignments as well as our more complex assignments. Each line of the file starts with a tag followed by a space followed by the data corresponding to the tag.

For example, Figure 2 shows a very simple configuration file that tests the classic hello world program written in Java. In that example, the first line that starts with the `RUN` tag specifies the script to use to evaluate the specification file. In the future, if we want to use a completely different specification format, we can change that first line to specify a different specification evaluation script.

```
RUN evaluate.sh
C javac edu/sjsu/CS001/Hello.java

T java edu.sjsu.CS001.Hello ben
X 0
O Hello ben

T java edu.sjsu.CS001.Hello
X 1
O USAGE: edu.sjsu.CS001.Hello name

HT java edu.sjsu.CS001.Hello "long name here!"
X 0
O Hello long name here!
```

Figure 2: Example CodEval specification.

The line that begins with a `C` tells CodEval how to compile the submission. The line that begins with a `T` tells CodEval how to run a test case for the submission and the following `X` and `O` lines specify the expected exit code and output for the test case. If there are multiple lines of output, multiple `O` lines would be specified. The last test case starts with `HT`, which indicates a hidden test case. Hidden test cases only indicate if the test passes or fails. Normal test cases will show the failing command and differences in expected results if the test case fails.

Table 1 lists all of the currently defined tags. Early in the first semester that we used CodEval we realized that we could also check the use of certain functions required by an assignment. The grader still needed to verify that the function was used correctly, but we could remind students to use these functions if they were missing from the assignment. For example, in the Operating Systems class the second assignment was focused on using `fork()`, so checking submissions for `fork()` helped remind students that were missing `fork()` in their submission. It also signaled to the grader a potential problem. Note, even if the submission was using `fork()`, the grader still needed to verify that the function was used correctly. Initially, we were using TCMD and grep to check for the use of functions, but the grep command was not as trivial as just `grep fork main.c` so we added the `CF` tag which allows us to specify `CF fork main.c`. The `CF`, `TCMD`, `IF`, and `OF` tags were added to CodEval after we had been using it for a while. In each case, we were able to add the tag and start using it in just a few minutes and with only a few lines of script. The CF command is implemented in four lines a `bash`.

| Tag | Meaning | Function |
| --- | --- | --- |
| `RUN` | Run Script | Specifies the script to use to evaluate the specification file. Defaults to `evaluate.sh`. |
| `Z` | Download Zip | Will be followed by zip files to download from Canvas to use when running the test cases. |
| `CF` | Check Function | Will be followed by a function name and a list of files to check to ensure that the function is used by one of those files. |
| `CMD/TCMD` | Run Command | Will be followed by a command to run. The TCMD will cause the evaluation to fail if the command exits with an error. |
| `CMP` | Compare | Will be followed by two files to compare. |
| `T/HT` | Test Case | Will be followed by the command to run to test the submission. |
| `I/IF` | Supply Input | Specifies the input for a test case. The IF version will read the input from a file. |
| `O/OF` | Check Output | Specifies the expected output for a test case. The OF version will read from a file. |
| `E` | Check Error | Specifies the expected error output for a test case. |
| `TO` | Timeout | Specifies the time limit in seconds for a test case to run. Defaults to 20 seconds. |
| `X` | Exit Code | Specifies the expected exit code for a test case. Defaults to zero. |

Table 1: CodEval specification tags.

*4.2. CodEval configuration*

The code eval configuration consists of two parts: configuration for Canvas and configuration for Docker. Figure 3 shows an example configuration. The SERVER section has the URL of the instance of the hosted Canvas instance. For San José State University, the URL is <https://sjsu.instructure.com>. The SERVER section also has the Canvas API token that corresponds to the user running CodEval. For security reasons, the example uses `XXXX` for the token.

The RUN section has the commands to run to prepare and run the evaluation of a submission. CodEval will substitute the words `SUBMISSIONS` and `EVALUATE` with the location of the downloaded submission and the command to run to evaluate the submission.

```
[SERVER]
url=https://sjsu.instructure.com
```

```
token=XXXX
[RUN]
precommand=
command=docker run -i -v SUBMISSIONS:/submissions jimg bash -c "cd /submissions; EVALUATE"
```

Figure 3: Example CodEval configuration file.

### 4.3. Running CodEval

CodEval takes the name of the course to evaluate as a runtime parameter. It will scan all the assignments with corresponding specification files in the Canvas Files section of the code and look for any submissions that do not have a CodEval comment. It will download and evaluate those submissions according to the CodEval specification. Since we want submissions to be evaluated within five minutes of submission, we use the `cron` service in Linux to run CodEval every five minutes.

### 4.4. Feedback Mechanism

CodEval performs a static and dynamic analysis. Static analysis verifies the presence of specific functions or commands and ensures the program compiles correctly. Ad-hoc static analysis can be performed using the `CMD` or `TCMD` commands. The dynamic analysis runs the compiled code against the test cases defined in the specification file and displays if the test case has passed or failed. If the test case has failed, it will display the command for the failed test case. We use the `diff` command to summarize failures due to unexpected output. Runtime analyzers such as `valgrind` [2], can also be used in dynamic analysis to check for correctness.

## 5   RESULTS

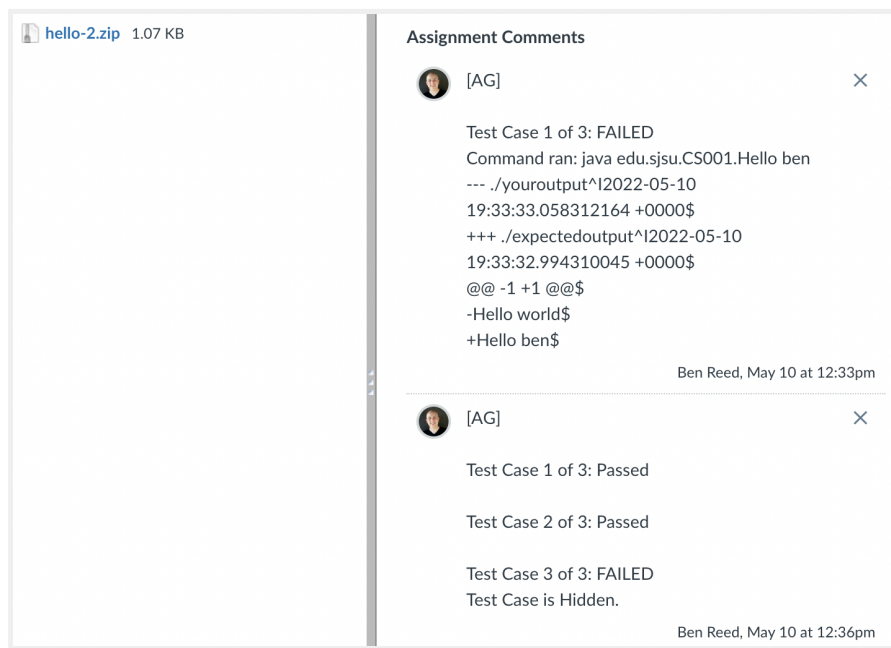

Figure 4: Example submission evaluation in Canvas.

Figure 4 shows an example of a submission evaluation for the Hello assignment specified in section 4.1. The submission is the classic hello world example which prints "hello world", but the assignment required the program to output "hello" followed by the name passed on the command line. CodEval caught the discrepancy and notified the student. The second submission printed "Hello ben", so it passed the first test. The submission also does the usage statement correctly for the second test, but the third test fails because the submission always prints "Hello ben".

## 5.1 Student evaluation

At the end of the semester we asked students for feedback on CodEval. We wanted to understand if they felt CodEval helped them with their assignments. A total of 28 students responded to the anonymous survey. Table 2 shows the results of the survey.

| Metric | Strongly Disagree | Somewhat Disagree | Neither Agree or Disagree | Somewhat Agree | Strongly Agree |
|---|---|---|---|---|---|
| Start assignment early | 0 | 6 | 12 | 6 | 4 |
| Improved code quality | 0 | 1 | 0 | 14 | 13 |
| Learn more from assignment | 0 | 4 | 6 | 12 | 6 |
| Improved overall success | 0 | 0 | 3 | 12 | 13 |

*Table 2: CodEval survey results.*

At the beginning of the semester, we hoped that because students were able to get immediate feedback on their assignments from CodEval that they would start their assignments earlier but did not seem to be the case. We did not notice a change in the desire for extensions or the absence of late submissions compared to previous semesters. As indicated in the student feedback only a few students saw CodEval as improving early starts to the assignments.

The results also indicate that students felt that CodEval helped them improve their submissions and succeed in the assignments. We also noted that no one complained about the grade that they received once the assignment was graded. That is despite the fact that CodEval doesn't actually grade the assignments and the grader must do their own evaluation of the correctness of the submission.

Apart from rating those four predefined dimensions, we also asked students to comment on how CodEval helped them and any improvements that they would like to see.

The students through their responses stated the following about the impact of CodEval:

1. immediate feedback was very valuable
2. they were able to catch errors that they wouldn't have otherwise
3. they were motivated to fix the errors in time
4. they were able to better understand and handle test cases
5. the tool helped their overall motivation
6. they were more confident about their submission
7. they could predict how well they did on an assignment before grades were awarded
8. helped them catch small bugs

The students suggested the following improvements:

1. better readability of the output results
2. faster response time on the grading
3. more detailed feedback on what was wrong with the code

4. hints on what can be done to fix errors
    5. removing hidden test cases

Note that improvements 2 and 3 were more about the grader than CodEval. We will be using improvements 1 and 4 to further improve CodEval. The students also complained about the hidden tests in person, so that was not an unexpected complaint. We try to emphasize that hidden tests are important to help students work out their own test cases and make sure they clearly understand the requirements of the assignment. We do not expect this complaint to go away ;)

## 5.2 Student evaluation

Four of the assignments were used in a previous instance of the class from two semesters earlier. We did not have CodEval back then and students did not get any feedback on their submissions until the assignments were fully graded. In Fall 2020 there were two different sections of the class. As we can see in Table 3, for the first three assignments the Spring 2022 average scores fell near or between the averages of the two sections. This suggests that for the early assignments the improvements that students felt in terms of success and code quality were more perceived than actual. However, the last assignment, which was a very difficult and complex assignment was clearly improved by CodEval.

In considering this comparison, it is important to note that we had different graders for Fall 2020 and Spring 2022. The graders for Fall 2020 also did not have the benefit of the testing of CodEval which is more rigorous than the ad-hoc scripts that graders in Fall 2020 use. The lack of rigor in testing may have inflated the scores in Fall 2020.

| Assignments | Difficulty | Fall 2020 Sec 2 | Fall 2020 Sec 3 | Spring 2022 Sec 1 |
|---|---|---|---|---|
| count the bits | easy | 94.94871795 | 91.02702703 | 92.6125 |
| count the bits faster | moderate | 88.76923077 | 93.75675676 | 87.6625 |
| flaky tests | difficult | 77.25641026 | 81.43243243 | 78.4125 |
| a big bag of strings | very difficult | 76.46153846 | 79.62162162 | 86.35 |

*Table 3: Comparing average assignment scores for Fall 2020 courses which did not use CodEval with Spring 2022 which did.*

## 6 CONCLUSIONS

As learning management systems such as Canvas continue to become more popular, we expect tools such as CodEval to become even more important for classes that involve programming assignment. Our goal was to create a system that provided timely feedback for students and relieved graders of the more tedious aspects of submission evaluation. We achieved both of these goals while still maintaining the submission and grading workflows of Canvas.

CodEval was simple to run and worked well throughout the semester. The design was simple enough that we were able to make enhancements we discovered quickly. We hope to expand the use of CodEval in other courses here at San José State University. We have also made CodEval open source. Instructors interested in using code eval can go to [15] for more information on how to use CodEval in their class.

Overall we were pleased with the performance of CodEval this semester. The students also gave positive feedback both in-person on in the survey as to the usefulness of CodEval. It also freed the grader of the burden of preparing and testing code prior to grading. CodEval also proved adaptable enough to address the requirements we had to evaluate the submissions of even our most difficult assignments.

## ACKNOWLEDGEMENTS

We would like to thank Instructure and the development community for providing invaluable direction in the use of the Canvas APIs. We also made extensive use of Python's 'canvasapi' package.